\documentclass[prd,preprint,superscriptaddress,showpacs,byrevtex]{revtex4}
\usepackage{epsfig}

\newcommand{\be}{\begin{equation}}
\newcommand{\ee}{\end{equation}}
\newcommand{\bea}{\begin{eqnarray}}
\newcommand{\eea}{\end{eqnarray}}

\begin{document}

\title{ Schwinger Mechanism for Gluon Pair Production
in the Presence of Arbitrary Time Dependent Chromo-Electric Field in Arbitrary Gauge }

\author{Gouranga C. Nayak} \email{gnayak@uic.edu}

\affiliation{ Department of Physics, University of Illinois, Chicago, IL 60607 USA }

\date{\today}

\begin{abstract}

We study non-perturbative gluon pair production from arbitrary time
dependent chromo-electric field $E^a(t)$ with arbitrary color index
$a$ =1,2,...8 via Schwinger mechanism in arbitrary covariant background gauge $\alpha$.
We show that the probability of non-perturbative gluon pair production per unit time
per unit volume per unit transverse momentum $\frac{dW}{d^4xd^2p_T}$ is
independent of gauge fixing parameter $\alpha$. Hence the result obtained
in the Fynman-'t Hooft gauge, $\alpha$=1, is the correct gauge
invariant and gauge parameter $\alpha$ independent result.

\end{abstract}
\maketitle

\newpage

An exact non-perturbative result for electron-positron
pair production from a constant electric field was obtained
by Schwinger in 1951 by using proper time method \cite{schw}. In QCD
this result depends on two independent casimir/gauge invariants $C_1=[E^aE^a]$
and $C_2=[d_{abc}E^aE^bE^c]^2$ with color indices $a,b,c$=1,2,...8 in SU(3)
\cite{gn,gn1}. Recently, using shift theorem \cite{shift}, we have extended this study
to arbitrary time dependent electric field $E(t)$ in QED \cite{qed} and to
arbitrary time dependent chromo-electric field $E^a(t)$ in QCD \cite{quark,gluon}.
This result crucially depends on the validity of the shift conjecture which is
not yet established.

In \cite{gluon} Schwinger mechanism for gluon pair production from arbitrary time dependent chromo-electric
field $E^a(t)$ was studied in the Feynman-t'hooft gauge $\alpha$=1.
In this paper we will extend this study to any arbitrary gauge fixing parameter $\alpha$.
We find that the result is gauge fixing parameter $\alpha$ independent.
Hence the result obtained in \cite{gluon} in Feynman-'t Hooft gauge, $\alpha$=1,
is the correct gauge invariant and gauge parameter $\alpha$ independent result.

The following result was obtained for the probability of gluon pair production
from arbitrary time dependent chromo-electric field $E^a(t)$ in $\alpha$=1 gauge
via Schwinger mechanism \cite{gluon}:
\bea
\frac{dW_{g (\bar g)}}{dt d^3x d^2p_T}~
=~\frac{1}{4 \pi^3} ~~ \sum_{j=1}^3 ~
~|g\Lambda_j(t)|~{\rm ln}[1~+~e^{-\frac{ \pi p_T^2}{|g\Lambda_j(t)|}}].
\label{1}
\eea
In the above equation
\bea
\Lambda_1^2=\frac{C_1(t)}{2}[1-{\rm cos}\theta(t)]; ~~~
\Lambda_{2,3}^2=\frac{C_1(t)}{2}[1+{\rm cos}(\frac{\pi}{3} \pm \theta(t))];~~~cos3\theta(t)=-1+6 C_2(t)/C_1^3(t) \nonumber \\
\label{lm}
\eea
\bea
{\rm where} ~~~~~~~~~~~~C_1(t)=[E^a(t)E^a(t)];~~~~~~~~~~~~~~ C_2(t)=[d_{abc}E^a(t)E^b(t)E^c(t)]^2
\label{cas}
\eea
are two independent time-dependent casimir/gauge invariants in SU(3).

We will present a proof of gauge fixing parameter $\alpha$ independence
of eq. (\ref{1}) in the following.

In the background field method of QCD \cite{thooft,abbott} the
gauge field is the sum of classical chromo-field $A_\mu^a$ and the quantum
gluon field $Q_\mu^a$. The non-abelian field tensor becomes
\bea
F_{\mu \nu}^a[A+Q]= \partial_\mu (A_\nu^a+Q_\nu^a) -\partial_\nu (A_\mu^a+Q_\mu^a) +gf^{abc} (A_\mu^b+Q_\mu^b) (A_\nu^c+Q_\nu^c).
\eea
The gauge field Lagrangian density is
\bea
{\cal L}_{gl}~=~-\frac{1}{4} F_{\mu \nu}^a[A+Q] F^{\mu \nu a}[A+Q]
~-\frac{1}{2\alpha} [D_\mu[A]Q^{\mu a}]^2
\label{total1}
\eea
where the second term in the right hand side is the gauge fixing term
which depends on the background field $A_\mu^a$ \cite{thooft,abbott}.
The covariant derivative is given by
\bea
D_\mu^{ab}[A]~=~\delta^{ab} \partial_\mu~+~gf^{abc}A_\mu^c.
\label{cov}
\eea
Keeping terms up to quadratic in $Q$ field (for gluon pair production)
we find from eq. (\ref{total1})
\bea
&&~\int d^4x {\cal L}= \frac{1}{2} \int d^4x~[-(D_\mu[A]Q_\nu^a)F^{\mu \nu a}[A] +
Q^{\mu a} M^{ab}_{\mu \nu}[A] Q^{\nu b} ] \nonumber \\
&& =\frac{1}{2} \int d^4x~[(D_\mu[A]F^{\mu \nu a}[A]) Q_\nu^a +
Q^{\mu a} M^{ab}_{\mu \nu}[A] Q^{\nu b} ]
\label{full}
\eea
where
\bea
M^{ab}_{\mu \nu}[A]~=~
g_{\mu \nu} [D_\rho(A)D^\rho(A)]^{ab}~-~2gf^{abc}F_{\mu \nu}^c[A]+(\frac{1}{\alpha} -1) (D_\mu[A] D_\nu[A])^{ab}
\label{mab}
\eea
with $g_{\mu \nu}=(1,-1,-1,-1)$.

The vacuum-to-vacuum transition amplitude for gluon is given by
\bea
<0|0>^A=\frac{Z[A]}{Z[0]}=
\frac{\int [dQ]e^{i\int d^4x [Q^{\mu a}M_{\mu \nu}^{ab}[A]Q^{\nu b}+(D_\mu [A] F^{\mu \nu a}[A])Q_\nu^a]
}}{\int [dQ] e^{i\int d^4xQ^{\mu a}M_{\mu \nu}^{ab}[0]Q^{\nu b}}}.
\label{vact}
\eea
To evaluate the path integration in eq. (\ref{vact}) we change the variable
\bea
{Q}_\mu^a(x)={Q'}_\mu^a(x)-\frac{1}{2}~\int d^4x' G_{\mu \nu}^{ab}(x,x')D_\lambda (x')F^{\lambda \nu b}(x')
\label{qp}
\eea
where we denote $D_\mu^{ab}(x)=D_\mu^{ab}[A](x)$ and
$F_{\mu \nu}^a(x)= F_{\mu \nu}^a[A](x)$. The Green's function is given by
(using Schwinger's notation \cite{schw})
\bea
G_{\mu \nu}^{ab}(x,x') = [<x|\frac{1}{M}|x'>]_{\mu \nu}^{ab}=[<x|\int_0^\infty ds~ e^{-sM }  |x'>]_{\mu \nu}^{ab}.
\label{green}
\eea
Under this change of variable we find from eq. (\ref{vact})
\bea
 <0|0>^A=\frac{Z[A]}{Z[0]}=e^{-i S_{\rm tad}} \times e^{iS^{(1)}}
\label{vact1}
\eea
where
\bea
S_{\rm tad} = \frac{1}{2}~\int d^4x \int d^4x' D_\mu(x) F^{\mu \lambda a}(x) G_{\lambda}^{ \nu ab}(x,x')
D^\sigma (x') F_{\sigma \nu b}(x')
\label{tad}
\eea
is the tadpole (or single gluon) effective action and
\bea
S^{(1)}=-i {\rm ln}[ \frac{{\rm Det^{-1/2}}M_{\mu \nu}^{ab}[A]}{{\rm Det^{-1/2}}M_{\mu \nu}^{ab}[0]}]
= \frac{i}{2}{\rm Tr}[ {\rm ln} M_\mu^{ \nu, ab}[A] -{\rm ln} M_\mu^{ \nu, ab}[0]]
\label{og}
\eea
is the one loop (or gluon pair) effective action.

The trace ${\rm Tr}$ is given by
\bea
{\rm Tr} {\cal O} = {\rm tr}_{\rm Lorentz } {\rm tr}_{\rm color}\int d^4x <x|{\cal O} |x>.
\label{tr}
\eea

We choose the arbitrary time-dependent chromo-electric field
$E^a(t)$ to be along the $z-$axis
(the beam direction) and work in the choice $A_3^a=0$ so that
\bea
A_\mu^a(x) = -\delta_{\mu 0} E^a(t) z
\label{gauge}
\eea
is non-vanishing. The color indices are arbitrary, $a$=1,2,...8.

It can be seen that the tadpole effective action $S_{\rm tad}$ in eq. (\ref{tad}) depends
on the gauge fixing parameter $\alpha$ via the Green's function $G_{\lambda}^{ \nu ab}(x,x')$.
However, we will show in the appendix that the tadpole effective action $S_{\rm tad}$
per unit time per unit volume and per unit transverse momentum ($\frac{dS_{\rm tad}}{d^4x d^2p_T}$)
is zero for any non-vanishing transverse momentum. Hence there is no tadpole contribution
to eq. (\ref{1}) and we will not consider it any more.

We write eq. (\ref{mab}) as
\bea
M^{ab}_{\mu \nu}[A]~=~M^{ab}_{\mu \nu,~\alpha=1}[A]~+~
\alpha'~(D_\mu[A] D_\nu[A])^{ab}
\label{m1abal}
\eea
where
\bea
\alpha' = (\frac{1}{\alpha}-1)
\label{alphap}
\eea
and
\bea
M^{ab}_{\mu \nu,~\alpha=1}[A]=g_{\mu \nu} [D_\rho(A)D^\rho(A)]^{ab}~-~2gf^{abc}F_{\mu \nu}^c[A].
\label{mabal}
\eea
Hence we find
\bea
&& {\rm Tr} {\rm ln} M_{\mu}^{ \nu, ab}[A] = {\rm Tr}{\rm ln} [M_{\mu,~\alpha=1}^\lambda [A]~\big \{\delta^\nu_\lambda
+\alpha'~{M^{-1~\sigma}_{\lambda,~\alpha=1}} [A]  ~(D_\sigma[A] D^\nu[A])\big \}]^{ab} \nonumber \\
&& = {\rm Tr}{\rm ln} [M_{\mu,~\alpha=1}^{\nu,~ab} [A]]~+~{\rm Tr}{\rm ln}[\delta_\mu^\nu \delta^{ab}
+\alpha'~[{M^{-1~\lambda}_{\mu,~\alpha=1}} [A]  ~D_\lambda[A] D^\nu[A]]^{ab}].
\label{mal1}
\eea
Since ${\rm Tr}{\rm ln} [M_{\mu,~\alpha=1}^{\nu,~ab} [A]]$ was studied in \cite{gluon} we will
evaluate the $\alpha' =(\frac{1}{\alpha}-1)$ dependent term in this paper.

The ghost determinant
is evaluated in \cite{gluon} where we have used $\alpha$=1. Since the ghost Lagrangian
density is $\alpha$ independent \cite{gluon}, we do not discuss ghost in this paper.
Whenever we mention $\alpha$ = 1 case in this paper, we assume that the ghost contribution
is included.

Using eq. (\ref{tr}) we find
\bea
&&{\rm Tr}{\rm ln}[\delta_\mu^\nu \delta^{ab}+\alpha'~[{M^{-1~\lambda}_{\mu,~\alpha=1}} [A]  ~D_\lambda[A] D^\nu[A]]^{ab}] = {\rm tr}_{\rm Lorentz } {\rm tr}_{\rm color} [\int d^2x_T <x_T| \int_{-\infty}^{+\infty} dt <t| \nonumber \\
&&  \int_{-\infty}^{+\infty} dz <z|~{\rm ln}[\delta_\mu^\nu+\alpha'~{M^{-1~\lambda}_{\mu,~\alpha=1}}  ~D_\lambda D^\nu ]|z>|t>|x_T>]^{ab}.
\label{mal2}
\eea
Using eq. (\ref{gauge}) in (\ref{cov}) we find
\bea
D_\mu^{ab} [A]=\partial_\mu \delta^{ab}-\delta_{\mu 0}~z~ig \Lambda^{ab}(t)
\label{d}
\eea
where
\bea
\Lambda^{ab}(t)=if^{abc}E^c(t).
\label{lamb}
\eea
Using the relation
\bea
[D_\mu[A],~D_\nu[A]]^{ab}=-gf^{abc}F_{\mu \nu}^c
\eea
and by using shift theorem \cite{shift} (by shifting $[z \rightarrow z + \frac{i}{ g \Lambda(t)}{\frac{d}{dt}}]^{ab}$~) we find from eq. (\ref{mal2})
\bea
&&{\rm Tr}{\rm ln}[\delta_\mu^\nu \delta^{ab}+\alpha'~[{M^{-1~\lambda}_{\mu,~\alpha=1}} [A]  ~D_\lambda[A] D^\nu[A]]^{ab}] = {\rm tr}_{\rm Lorentz } {\rm tr}_{\rm color} [\int d^2x_T <x_T| \int_{-\infty}^{+\infty} dt <t| \nonumber \\
&&  \int_{-\infty}^{+\infty} dz <z+\frac{i}{g\Lambda(t)}\frac{d}{dt}|~{\rm ln}[\delta_\mu^\nu+\alpha'~{(M')^{-1~\lambda}_{\mu,~\alpha=1}}   ~D'_\lambda D'^\nu ]|z+\frac{i}{g\Lambda(t)}\frac{d}{dt}>|t>|x_T>]^{ab} \nonumber \\
&& = {\rm tr}_{\rm Lorentz } {\rm tr}_{\rm color} [\int d^2x_T <x_T| \int_{-\infty}^{+\infty} dt <t| \nonumber \\
&&  \int_{-\infty}^{+\infty} dz <z+\frac{i}{g\Lambda(t)}\frac{d}{dt}|~{\rm ln}[\delta_\mu^\nu+\alpha' ~D'_\mu \frac{1}{(D')^2} D'^\nu ]|z+\frac{i}{g\Lambda(t)}\frac{d}{dt}>|t>|x_T>]^{ab}
\label{mal3}
\eea
where
\bea
{D'}_\mu^{ab} [A] = (1-\delta_{\mu 0})~\delta^{ab}~\partial_{\mu } -
\delta_{\mu 0}~z~ig \Lambda^{ab}(t)
\label{dp}
\eea
($\mu$ is not summed) and
\bea
{M'}^{ab}_{\mu \nu,~\alpha=1}[A]=g_{\mu \nu} [D'_\rho(A){D'}^\rho(A)]^{ab}~-~2gf^{abc}F_{\mu \nu}^c[A].
\label{m1ab}
\eea
It has to be remembered that the $z$ integration must be performed from
-$\infty$ to +$\infty$ for the shift theorem \cite{shift} to be applicable.

Expanding the Logarithm in eq. (\ref{mal3}) we find
\bea
&& {\rm ln}[\delta_\mu^\nu+\alpha' ~D'_\mu \frac{1}{(D')^2} D'^\nu ]^{ab}
=\alpha' [D'_{\mu} \frac{1}{(D')^2} D'^{\nu} ]^{ab} \nonumber \\
&&~-~
\frac{{\alpha'}^2}{2}~ [ D'_{\mu} \frac{1}{(D')^2} D'^{\nu} ]^{ab}
~+~
\frac{{\alpha'}^3}{3}~ [ D'_{\mu} \frac{1}{(D')^2} D'^{\nu} ]^{ab}
~-~ \frac{{\alpha'}^4}{4}~ [ D'_{\mu}
\frac{1}{(D')^2} D'^{\nu} ]^{ab}  \nonumber \\
&& ~+~ \frac{{\alpha'}^5}{5}~ [ D'_{\mu}
\frac{1}{(D')^2} D'^{\nu} ]^{ab}
~-~.........
\label{trmab6}
\eea
Summing the series we obtain
\bea
{\rm ln}[\delta_\mu^\nu+\alpha' ~D'_\mu \frac{1}{(D')^2} D'^\nu ]^{ab}
= {\rm ln} (1+\alpha')~  [ D'_{\mu} \frac{1}{(D')^2} D'^{\nu} ]^{ab}.
\label{trans}
\eea
Using the cyclic properties of the trace
(${\rm Tr} [D'_\mu \frac{1}{(D')^2} D'^\nu ]^{ab}
={\rm Tr} [D'^\nu D'_\mu \frac{1}{(D')^2} ]^{ab}$) and using eq. (\ref{trans})
we find from eq. (\ref{mal3})
\bea
&&{\rm Tr}{\rm ln}[\delta^\mu_\nu \delta^{ab}+\alpha'~[{M^{-1~\lambda}_{\mu,~\alpha=1}} [A]  ~D_\lambda[A] D^\nu[A]]^{ab}=
{\rm tr}_{\rm Lorentz } {\rm tr}_{\rm color} [\int d^2x_T <x_T| \int_{-\infty}^{+\infty} dt <t| \nonumber \\
&&  \int_{-\infty}^{+\infty} dz <z+\frac{i}{g\Lambda(t)}\frac{d}{dt}|~D'_\mu \frac{1}{(D')^2} D'^\nu ~{\rm ln}[1+\alpha']~|z+\frac{i}{g\Lambda(t)}\frac{d}{dt}>|t>|x_T>]^{ab} \nonumber \\
&& ={\rm tr}_{\rm color} [\int d^2x_T <x_T| \int_{-\infty}^{+\infty} dt <t|  \int_{-\infty}^{+\infty} dz <z+\frac{i}{g\Lambda(t)}\frac{d}{dt}| ~{\rm ln}[1+\alpha']~|z+\frac{i}{g\Lambda(t)}\frac{d}{dt}> \nonumber \\
&& |t>|x_T>]^{ab} ~=~8~{\rm ln}[1+\alpha'] \int d^4x \int d^4p~=~-8~{\rm ln}(\alpha) \int d^4x \int d^4p
\label{mal4}
\eea
where we have used $\alpha' = (\frac{1}{\alpha}-1$) from eq. (\ref{alphap}).
Using eq. (\ref{mal4}) in (\ref{mal1}) we find
\bea
{\rm Tr} {\rm ln} M_{\mu}^{ \nu, ab}[A] = {\rm Tr}{\rm ln} [M_{\mu,~\alpha=1}^{\nu,~ab} [A]]
-8 ~{\rm ln}(\alpha) \int d^4x \int d^4p.
\label{mal5}
\eea
Similarly for the free part we get
\bea
{\rm Tr} {\rm ln} M_{\mu}^{ \nu, ab}[0] = {\rm Tr}{\rm ln} [M_{\mu,~\alpha=1}^{\nu,~ab} [0]]
-8 ~{\rm ln}(\alpha) \int d^4x \int d^4p.
\label{mal6}
\eea
Using eqs. (\ref{mal5}) and (\ref{mal6}) in eq. (\ref{og}) we find
\bea
S^{(1)} = \frac{i}{2}{\rm Tr}[ {\rm ln} M_\mu^{ \nu, ab}[A] -{\rm ln} M_\mu^{ \nu, ab}[0]]= \frac{i}{2}{\rm Tr}[ {\rm ln} M_{\mu,~\alpha=1}^{ \nu, ab}[A] -{\rm ln} M_{\mu,~\alpha=1}^{ \nu, ab}[0]]
\label{vac4}
\eea
where the gauge parameter $\alpha$ dependence exactly cancelled from the interacting and free part.
The imaginary part of this effective action $S^{(1)}$ gives real gluon pair production result
(eq. (\ref{1})) \cite{gluon}.
Hence we find that the non-perturbative result for gluon pair production from arbitrary $E^a(t)$ via Schwinger
mechanism is independent of arbitrary gauge parameter $\alpha$ which is used in the gauge fixing term in the
background field method of QCD.

To conclude we have studied the Schwinger mechanism
for gluon pair production in the presence of arbitrary time-dependent chromo-electric field $E^a(t)$
in arbitrary covariant background gauge $\alpha$
with arbitrary color index $a$=1,2,..8 by directly evaluating the path integral.
We have found that the exact result for
non-perturbative gluon pair production from arbitrary $E^a(t)$ via Schwinger mechanism
is independent of arbitrary gauge parameter
$\alpha$ which is used in the gauge fixing term in the background field
method of QCD. We find that the non-perturbative gluon pair production result
from arbitray $E^a(t)$ via Schwinger mechanism
is both gauge invariant and gauge parameter $\alpha$ independent. 
Gluon production from classical chromo
field may be relevant to study production of quark-gluon plasma at RHIC and LHC.

\acknowledgments
I thank Peter van Nieuwenhuizen for useful discussions. This work was
supported in part by the U.S. Department of Energy under Grant no. DE-FG02-01ER41195.

\appendix

\section{ }

By using eq. (\ref{gauge}) in (\ref{cov}) we find
\bea
&& i(D^\mu[A])^{ab}=\delta^{\mu 0} (\delta^{ab}{\hat p}_0 -g\Lambda^{ab}(t)z)+ \delta^{ab}\delta^{\mu 1} \hat{p}_x +\delta^{ab} \delta^{\mu 2} {\hat p}_y+\delta^{ab} \delta^{\mu 3} \hat{p}_z  \nonumber \\
&& =\delta^{\mu 0} (\delta^{ab}{\hat p}_0 -g\Lambda^{ab}(t)z)+ \delta^{ab}\delta^{\mu T} \hat{p}_T + \delta^{ab}\delta^{\mu 3 } \hat{p}_z
\label{dmu}
\eea
where $\Lambda^{ab}(t)$ is given by eq. (\ref{lamb}).
From eqs. (\ref{mab}) and (\ref{green}) we obtain
\bea
&& G_\mu^{\nu, ab}(x,x')
= \int_0^\infty ds[<x| \int_0^\infty ds ~\nonumber \\
&& e^{-s( \delta_\mu^\nu [-({\hat p}_0 -
g\Lambda(t) z)^2 +\hat{p}_z^2 +\hat{p}_T^2]-~2g \Lambda(t){\hat F}_\mu^\nu
-(\frac{1}{\alpha}-1)(\delta_{\mu 0} ({\hat p}_0 -g\Lambda(t)z)- \delta_{\mu T} \hat{p}_T
- \delta_{\mu 3} \hat{p}_z)(\delta^{\nu 0} ({\hat p}_0 -g\Lambda(t)z)+ \delta^{\nu T} \hat{p}_T +\delta^{\nu 3} \hat{p}_z)
 )} \nonumber \\
&&  |x'>]^{ab}.
\label{gr}
\eea
where
\bea
{\hat F}_\mu^\nu=\delta_{\mu 3} \delta_{\nu 0}+\delta_{\mu 0} \delta_{\nu 3}.
\label{fm}
\eea
Using eqs. (\ref{gr}) and (\ref{gauge}) in (\ref{tad}) we find the tadpole effective action
\bea
&& S_{\rm tad} = \frac{1}{2}\int d^2x_T ~d^2x'_T~  dz~ dt~ dz'~ dt' ~\int_0^\infty ds~ \frac{dE^a(t)}{dt}
[<x_T|  <t|<z| \nonumber \\
&& e^{-s( \delta_\mu^\nu [-({\hat p}_0 -
g\Lambda(t) z)^2 +\hat{p}_z^2 +\hat{p}_T^2]-~2g \Lambda(t){\hat F}_\mu^\nu
-(\frac{1}{\alpha}-1)(\delta_{\mu 0} ({\hat p}_0 -g\Lambda(t)z)- \delta_{\mu T} \hat{p}_T - \delta_{\mu 3} \hat{p}_z)(\delta^{\nu 0} ({\hat p}_0 -g\Lambda(t)z)+ \delta^{\nu T} \hat{p}_T +\delta^{\nu 3} \hat{p}_z))} \nonumber \\
&&  |t'>|z'> x'_T>]^{ab} \frac{dE^b(t')}{dt'}.
\label{tad2}
\eea
Inserting complete set of $|p_T>$ states (by using $\int d^2p_T |p_T><p_T|=1$) we find
\bea
&& S_{\rm tad} =\frac{1}{2} \int d^2x_T ~d^2x'_T~  dz~ dt~ dz'~ dt' ~ d^2p_T~\int_0^\infty ds~ \frac{dE^a(t)}{dt}
[<x_T|p_T>  <t|<z| \nonumber \\
&& e^{-s( \delta_\mu^\nu [-({\hat p}_0 -
g\Lambda(t) z)^2 +\hat{p}_z^2 +p_T^2]-~2g \Lambda(t){\hat F}_\mu^\nu -(\frac{1}{\alpha}-1)(\delta_{\mu 0} ({\hat p}_0 -g\Lambda(t)z)- \delta_{\mu T} p_T - \delta_{\mu 3} \hat{p}_z)(\delta^{\nu 0} ({\hat p}_0 -g\Lambda(t)z)+ \delta^{\nu T} p_T +\delta^{\nu 3} \hat{p}_z)
)} \nonumber \\
&& |t'>|z'> <p_T|x'_T>]_3^{3,~ab}
\frac{dE^b(t')}{dt'}
\label{tad1}
\eea
where $_3^3$ means the $\mu=3$ and $\nu=3$ component of the Lorentz matrix.
Using $<q|p>=\frac{1}{\sqrt{2\pi}} e^{iqp}$ we obtain
\bea
&& S_{\rm tad} = \frac{1}{2(2\pi)^2} \int d^2x_T ~d^2x'_T ~ dz ~dt ~dz' ~dt'~  d^2p_T~ \int_0^\infty ds~\frac{dE^a(t)}{dt}
[e^{ix_T \cdot p_T}   <t|<z| \nonumber \\
&&e^{-s( \delta_\mu^\nu [-({\hat p}_0 -
g\Lambda(t) z)^2 +\hat{p}_z^2 +p_T^2]-~2g \Lambda(t){\hat F}_\mu^\nu -(\frac{1}{\alpha}-1)(\delta_{\mu 0} ({\hat p}_0 -g\Lambda(t)z)- \delta_{\mu T} p_T - \delta_{\mu 3} \hat{p}_z)(\delta^{\nu 0} ({\hat p}_0 -g\Lambda(t)z)+ \delta^{\nu T} p_T +\delta^{\nu 3} \hat{p}_z)
 )} \nonumber \\
&&e^{-ix'_T \cdot p_T}  |z'>|t'>]_3^{3 ab}
\frac{dE^b(t')}{dt'}.
\label{tad3}
\eea
Integrating over $x'_T$ (by using $\int d^2x'_T e^{-ix'_T \cdot p_T} = (2 \pi)^2 ~\delta^{(2)}(\vec{p}_T))$
we find
\bea
&& \frac{dS_{\rm tad}}{dt d^3xd^2p_T} =\frac{\delta^{(2)}(\vec{p_T})}{2}~ \int dt'~\int dz'  ~\int_0^\infty ds~\frac{dE^a(t)}{dt}
[<t|<z| \nonumber \\
&& e^{-s( \delta_\mu^\nu [({\hat p}_0 -
g\Lambda(t) z)^2 +\hat{p}_z^2 +p_T^2]-~2g \Lambda(t){\hat F}_\mu^\nu -(\frac{1}{\alpha}-1)(\delta_{\mu 0} ({\hat p}_0 -g\Lambda(t)z)- \delta_{\mu T} p_T - \delta_{\mu 3} \hat{p}_z)(\delta^{\nu 0} ({\hat p}_0 -g\Lambda(t)z)+ \delta^{\nu T} p_T +\delta^{\nu 3} \hat{p}_z)
 )} \nonumber \\
&& |z'>|t'>]_3^{3 ab} \frac{dE^b(t')}{dt'}.
\label{tad4}
\eea
which has a $\delta^{(2)}(\vec{p}_T)$ distribution. Hence we find for any non-vanishing $p_T$
\bea
\frac{dS_{\rm tad}}{d^4xd^2p_T} = 0.
\label{ftad}
\eea
Hence the tadpole (or single gluon) effective action do not
contribute to eq. (\ref{1}) of the non-perturbative gluon (pair) production
rate $\frac{dW}{d^4xd^2p_T}$ via Schwinger mechanism.

\end{document}